\documentclass[prb,floatfix,aps,twocolumn,showpacs,groupedaddress]{revtex4-1}
\usepackage[dvips]{graphics}
\usepackage[english]{babel}
\usepackage{graphicx}
\usepackage[utf8]{inputenc}
\usepackage{amssymb}
\usepackage{epstopdf}
\usepackage{bm}
\usepackage{dcolumn}
\usepackage{epsfig}
\usepackage{latexsym}
\usepackage{amsmath}
\usepackage{color}
\usepackage{array}
\usepackage{enumerate,dsfont}
\usepackage{mathtools}

\def\XXint#1#2#3{{\setbox0=\hbox{$#1{#2#3}{\int}$ }
\vcenter{\hbox{$#2#3$ }}\kern-.5\wd0}}

\definecolor{mydarkgreen}{rgb}{0,0.5,0}

\usepackage{braket}

\begin{document}
\title{Boundary spin polarization as  robust signature of topological phase transition in Majorana nanowires}
\author{Marcel Serina, Daniel Loss, and Jelena Klinovaja}
\affiliation{Department of Physics, University of Basel, Klingelbergstrasse 82,
CH-4056 Basel, Switzerland}
\date{\today}
\begin{abstract}
We show that the boundary charge and spin can be used as alternative signatures of the topological phase transition in topological models such as semiconducting nanowires with strong Rashba spin-orbit interaction in the presence of a magnetic field and in proximity to an $s$-wave superconductor. We identify signatures of the topological phase transition that do not rely on the presence of Majorana zero-energy modes and, thus, can serve as independent probes of topological properties. The boundary spin component along the magnetic field, obtained by summing contributions from all states below the Fermi level, has a pronounced peak at the topological  phase transition point. Generally, such signatures can be observed at boundaries between topological and trivial sections in nanowires and are stable against disorder.
\end{abstract}
\maketitle

\section{Introduction}
Topological models have attracted a lot of attention in recent years. One of the first topological systems proposed about fourty years ago is the  Su-Schrieffer-Heeger (SSH) model~\cite{SSHSu},  describing properties of one-dimensional dimerized polymers. In this spinless model, a nondegenerate fermionic zero-mode, localized at a domain wall, is associated with a well-defined  half-integer boundary charge \cite{FractChargeSu, FracSolitGoldstone}. The same results were first predicted in a continuum model proposed by Jackiw and Rebbi \cite{HalfFermJackiw}. The half-integer value of the boundary charge in these models is protected by the chiral symmetry. If this symmetry is broken, the value of the boundary charge can deviate from $e/2$~\cite{JackiwSchrieffSolit,SuSchrieffFractCharg}. Importantly, however, in the topological regime, there is always a boundary charge (independent of the presence of bound states) at the domain wall as was shown in several extensions of the SSH model~\cite{RiceMeleDiatPolym,JackiwSemenfDiatPolym,KivelsonSolit}. Recently, the concept of the fractional boundary charge in topological SSH models was revisited, aiming at different systems that could be realized in modern experimental settings \cite{QiHughesZhangHall,GoldmanSatijaTRTopoIns,GangaTrifunovicEndStatWire, KrausTopoQuasicryst,FractFermMajKlinovaja, BudichFractTopo,ZhuLiFractTopoOptSuperlatt, GrusdtTopoBoseHubbard,NonAbelFracFermKlinovaja, MadsenTopoEqCrystQuasicryst,  PoshakinskiyTopoPhotCryst, TranspSignFractFermRainis, ParafNWireBudleKlinovaja, WakatsukiFermFractToMaj, MarraFractChargSuperlatt, KlinovajaFermMajBound,MiertExcessCharge,SSHLongRangeHop,Chamon} 
and even in higher dimensions~\cite{SeradjehFracTRSymm, RueggFractChargeKagome, SzumniakEdgeStates}. Motivated by these studies on boundary charges we would like to go a step further and focus in this work on boundary spins. In particular, we want to study the behavior of boundary spins in- and outside
the  topological phase and demonstrate that the total moment of spins close to the boundary can be used as signature for the topological phase transition.

Currently, Majorana fermions (MFs), proposed as a real-field solution of the Dirac equation and thus being its own antiparticle~\cite{EttMaj}, attract the most attention among the known bound states in topological systems. With the rapidly growing interest in  topological properties of condensed matter systems~\cite{TopoVolkov, BernevigScience, MolenkampScience1, KaneMele, MolenkampScience2, RevModPhysKane, BookTopoSCBernevig},  MFs  were proposed to be present in various theoretical and experimental setups~\cite{MajFSurfTopoInsFu,TopoPhasesSato, DSarmaNanowMaj,OppenNanowMaj,MajFermAlicea,MajMultibRashbaPotter, MajoranaCarbNantbKlinovaja, AndrIntoMajChevallier,  MajMagnAdatYazdani, OppenShibaTh, MajRKKYKlinovaja, MajMagMomSCSimon,SelfOrgTopoStateVazifeh,FloqMajThakurathi, MajGeSiHoleMaier, DynDetTPTSetiawan, WirelMajZutic}. The most promising ones among them being chains of magnetic adatoms on  superconducting surfaces~\cite{YazdaniShiba,OppenShibaExp,MeyerShiba} and  semiconducting nanowires (NWs) with sizeable Rashba spin-orbit interaction (SOI) in the presence of a magnetic field and proximity-induced superconductivity~\cite{KouwMaj,ExpMajDas, ExpMajRokhinson,ExpMajMarcus, MarcusMaj}. Majorana fermions can be used as building blocks for  topological quantum computing~\cite{FaultTolQCompKitaev,RevModPhysQComp} and can be combined with spin qubits in quantum dots into hybrid architectures~\cite{MajDotFlensberg,MajDotLiu,MajLeakQDEgues,MajHybQub,MajDotExpDeng,DecBndStateRicco, QuantPhaseTrMajDessotti,MajSurfCodeEgger, TopoPhiZeroSchrade,dcJosepQDXu,MajQDPrada}.

\begin{figure}[t!]
\renewcommand\figurename{Figure} 
\centering
\includegraphics[width=\columnwidth]{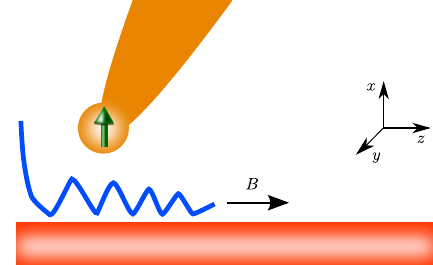}
\caption{\label{NVMajoranaFig} Our setup consists of a  semiconducting NW (red bar) with Rashba SOI and with proximity-induced superconductivity  due to coupling to a bulk $s$-wave superconductor (not shown). The SOI vector $\alpha$ points along the $y$ direction. A magnetic field $\bf B$ applied along the NW axis is used to tune the system in and out of the topological phase. The boundary spin accumulates at the edges of the NW (blue curve) and can be used to identify the topological phase transition point. The magnetic signal can be measured by using {\it e.g.} NV centers (green arrow) on a tip (orange). }
\end{figure}

Most of the studies~\cite{ChargeTransMajFu,NSMajAguado,MajNWNormLeadSerra,
MajorSCCorrelRainis,MajorTranspRainis,CoulMajJuncEgger,
MultitermNetwMajSchmidt,HeliLiqMBSDolcini,DisordSCSarma,SpinReversSzumniak,MajQDPrada} so far focused  on the transport properties of such NWs in the topological regime or on properties of  MFs themselves and their dependence on physical parameters.
Also, there has been substantial interest recently in the investigation of the spin polarization of Andreev and Majorana bound states \cite{MajSpinPolSimon,MajSpinPolBena}. However, it has been pointed out that  great care must be taken when identifying topological phases from the presence of quasiparticle states inside the superconducting gap \cite{MajSpinTewari,AndrBSvsMajBSSarma,TwoTermChargeTunnTewari}. Thus, it is most desirable to have additional signatures  available (besides MFs) that would allow one to identify the topological phase transition. 
Recent works, which analyzed the bulk signatures of the topological transition, focused either on the spinless Kitaev model~\cite{BulkSignKitaevModelChan} or studied finite-size scaling of the ground state energy in a generic conformal field theory approach for each of the symmetry classes~\cite{FinSizeScalCFTKamenev}. 
In this work, we would like to investigate the experimentally most relevant model of  Rashba NWs and to provide relevant quantities accessible by state-of-the-art measurements. In contrast to  aformentioned works, we also focus on local boundary effects and consider here  different aspects of topological phases in one-dimensional systems, namely non-transport signatures of the topological phase transition in the \textit{bulk} states, or, more precisely, in the boundary charge and boundary spin to which all occupied states close to the Fermi level contribute. 

The paper is organized as follows. In Sec. II, we introduce the Rashba NW setup, which is modeled and analyzed in a tight-binding description by means of exact diagonalization, where all four components of the quasiparticle wavefunctions, needed for calculating the observables of interest, are obtained. In Sec. III, we focus on the boundary spin in the  topological and trivial phases and find pronounced signatures in the spin component along the magnetic field direction, which allows us to identify the topological phase transition point. We summarize our results in Sec. IV.

\section{Model}
We investigate the system composed of a semiconducting NW with strong Rashba SOI in the proximity to a bulk $s$-wave superconductor in the presence of a magnetic field $B$ applied along the NW axis along $z$ direction, see Fig. \ref{NVMajoranaFig}. The SOI vector $\alpha$ points along the $y$ direction. The $B$-field results in the Zeeman energy  $\Delta_Z = g \mu_B B/2$, where $g$ is $g$-factor and $\mu_B$ the Bohr magneton. The corresponding tight-binding Hamiltonian is written as~\cite{MajorTranspRainis} 
\begin{align}\label{Hamiltonian}
	&H =  -t\sum_{\left\langle jj^{\prime}\right\rangle\sigma}\left(c_{j\sigma}^\dagger
	c_{j^{\prime}\sigma}+\mathrm{H.c.}\right)+
	\left(2t-\mu\right)\sum_{j\sigma}c_{j\sigma}^{\dagger}c_{j\sigma}\nonumber\\ & + \Delta_{Z}\sum_{j\sigma\sigma'}c_{j\sigma}^{\dagger}\sigma^{\sigma\sigma^\prime}_z c_{j\sigma^{\prime}} 
	 +i\sum_{\left\langle jj^{\prime}\right\rangle\sigma\sigma'}\left(\alpha c_{j\sigma}^{\dagger}\sigma^{\sigma\sigma^\prime}_y c_{j^{\prime}\sigma^\prime} +\mathrm{H.c.}\right)\nonumber\\
	&+\Delta_{SC}\sum_{j}\left(c_{j\uparrow}^{\dagger}c_{j\downarrow}^{\dagger}+\mathrm{H.c.}\right),
\end{align} 
where the creation operator $c_{i\sigma}^\dagger$ creates an electron with spin $\sigma=\uparrow,\downarrow$ at site $j$ of a chain consisting of $N$ sites with lattice constant $a$. In the first and fourth terms, the summation runs only over neighbouring sites $j$ and $j'$.
Here, $t$ denotes a nearest-neighbour hopping matrix element, $\mu$ is the chemical potential, and $\Delta_{SC}$ denotes the superconducting gap induced by proximity to the bulk $s$-wave superconductor.  We note that in our model $\mu=0$ corresponds the chemical potential being tuned to the SOI energy, which is defined here as $E_{SO}=\alpha^2/t$. For the rest of the paper we fix $t=1$ and use it as an energy scale. 
The system is in the topological phase hosting zero-energy MFs at the nanowire ends if $\Delta_Z>\Delta_Z^c$, where $\Delta_Z^c=\sqrt{\mu^2+\Delta_{SC}^2}$ \cite{DSarmaNanowMaj,OppenNanowMaj}. To study the topological phase transition in semiconducting NWs, we focus on the experimentally most relevant strong SOI regime, $E_{SO}\gg \Delta_{SC},\Delta_{Z}$~\cite{MarcusMaj,KouwMaj}.

\begin{figure}[t!]
    \renewcommand\figurename{Figure} 
\centering
\includegraphics[width=\columnwidth]{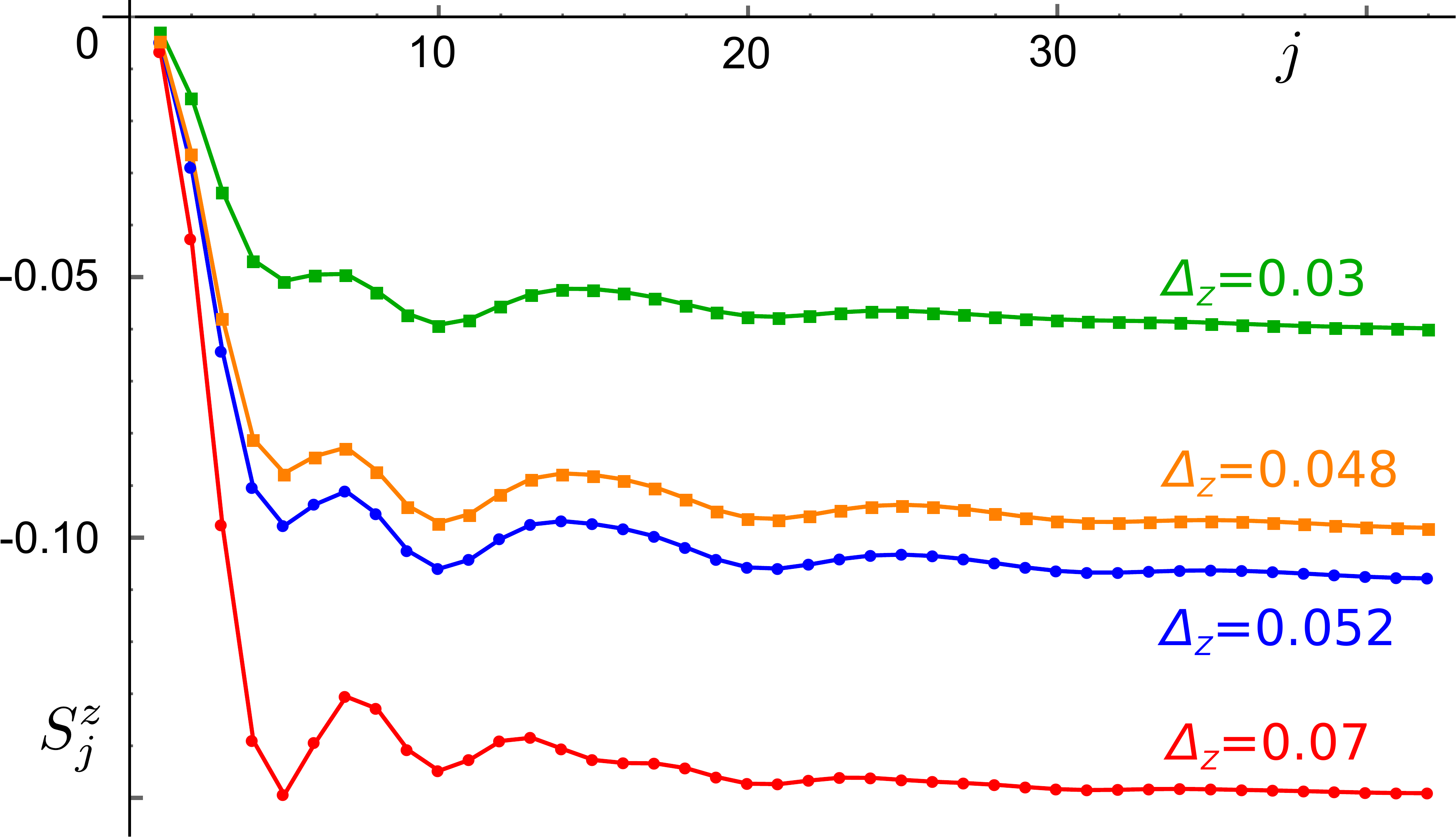}
  \caption{\label{ZSpinDensPlot} The spin density $S^z_j$ (projecton along $\bf B$-field) as a function of site $j$. Away from the NW ends,  $S^z_j$ is constant and given by $S^z_0$ determined solely by bulk properties of the system. However, close to NW ends, $S^z_j$  oscillates in all parameter regimes and there is no qualitative difference between trivial (indicated by squares) and topological (indicated by dots) phase. In all plots, the lines are guides to the eye. However, there is a quantitative shift in the amplitude of the first oscillations, as the system is driven through the topological phase transition.  We consider the system being deep in (red, $\Delta_Z=0.07$) and out of (green, $\Delta_Z=0.03$) the topological phase as well as close to the topological phase transition in (blue, $\Delta_Z=0.048$) and out of (orange, $\Delta_Z=0.052$) the topological phase. The parameters are fixed to $N=150$, $\mu=0$, $\alpha=0.3$, and $\Delta_{SC}=0.05$. This choice of parameters corresponds to typical values observed in experiments with nanowires such as: $m=0.015m_e$, $v_F=0.8\times 10^6$ m/s, $E_{SO}=0.9$ meV, and $\Delta_{SC}=0.5$ meV (with $a=15$ nm, $t=10$ meV).
}
\end{figure}

By diagonalizing numerically the Hamiltonian $H$ [see Eq.~(\ref{Hamiltonian})], one can determine the energy spectrum $E_n$. In addition, one also finds the operators $\psi_n=\sum_j \left(u^*_{\uparrow n j} c_{\uparrow j} + u^*_{\downarrow n j}c_{\downarrow j}+ v^*_{\uparrow n j}c^\dagger_{\uparrow j} +v^*_{\downarrow n j}c^\dagger_{\downarrow j}\right)$, corresponding to annihilation operators for each of these $n=4N$ states. Due to particle-hole symmetry, all states  appear in pairs, {\it i.e.} if $E_n$ is a solution, then so is $-E_n$. In what follows, we will focus on non-positive energy states. To characterize local bulk properties, we define the local charge $\rho_j$ and the local spin densities $S^{x,y,z}_j$  at each site as
\begin{subequations}\label{Densities}
\begin{align}
\rho_j=& \sum_{E_n<0;\sigma}\left(|u_{\sigma n j}|^2-|v_{\sigma n j}|^2\right),\\ 
S^{z}_j=& \sum_{E_n<0;\sigma}\sigma\left(|u_{\sigma n j}|^2-\sigma|v_{\sigma n j}|^2\right),\\
S^{x}_j=& \sum_{E_n<0;\sigma}\left(u_{\sigma n j}u^{*}_{\bar{\sigma} n j}-v_{\sigma n j}v^{*}_{\bar{\sigma} n j}\right),\\
S^{y}_j=& i\sum_{E_n<0;\sigma}\left(\sigma u_{\sigma n j}u^{*}_{\bar{\sigma} n j}-\sigma v_{\sigma n j}v^{*}_{\bar{\sigma} n j}\right),
\end{align}
\end{subequations}

where the index $\sigma=1$ ($\bar1$) corresponds also to spin up (down) states defined above, see Fig. \ref{ZSpinDensPlot}.
For zero-energy MF wavefunctions one can show that  $u^*_{\uparrow n j}=v_{\uparrow n j}$ and $u^*_{\downarrow n j}=v_{\downarrow n j}$. Thus, the MF charge and spin densities are exactly zero ~\cite{CompMajKlinovaja} and they do not contribute to Eq.~(\ref{Densities}). For this reason,  we take in our definition
 only bulk states with negative energies into account.  In addition, in our model, the Hamiltonian is real, so all the eigenvectors  can also be chosen to be real. As a consequence, we find that  $S^y$ is identically zero for all configurations considered below.

Away for the NW ends, both spin and charge densities are constant as expected in a translationally invariant system, see Fig. \ref{ZSpinDensPlot}. However, close to the NW end, these quantities oscillate around their bulk values $\rho_0$ and $S^{x,z}_{0}$ determined as the value at the middle of the NW $\rho_0=\rho_{j=\left[\frac{N}{2}\right]}$ and $S^{x,z}_{0}=S^{x,z}_{j=\left[\frac{N}{2}\right]}$, where $\left[M\right]$ denotes the integer part of $M$. Here, we assume that the NW is long enough such that these oscillations decay in the middle of the NW. In the strong SOI regime~\cite{SpinReversSzumniak,CompMajKlinovaja}, there are two lengthscales associated with bulk gaps at exterior branches $\xi_{e}/a=2\alpha/\Delta_{SC}$  and interior branches $\xi_{i}/a=2 \alpha/\left|\Delta_{SC}-\Delta_Z\right|$. In what follows, we work in the regime in which the NW length $L$ is much longer than both $\xi_{e}$ and $\xi_{i}$, see Fig. \ref{ZSpinDensPlot}.

Our main interest are boundary effects.  As a result, for further convenience~\cite{EdgeChargeLoss}, we define the left and right boundary charge and spin as
\begin{align}
&\tilde{\rho}_{Lm}= \sum^{m}_{j=1}\left(\rho_j-\rho_0\right),\\ &\tilde{S}^{x,z}_{Lm}= \sum^{m}_{j=1}\left( S^{x,z}_j-S^{x,z}_0\right), \label{EdgeDens}
\\ &\tilde{\rho}_{Rm}= \sum^{N}_{j=N-m}\left(\rho_j-\rho_0\right),\\ &\tilde{S}^{x,z}_{Rm}= \sum^{N}_{j=N-m}\left( S^{x,z}_j-S^{x,z}_0\right).
\end{align}
First, we subtract from charge and spin densities their bulk values. Second, we sum densities over $m$ sites at the left or right edge to define the right and left boundary charge or spin. Our system is symmetric with respect to the middle of the NW, so right and left boundary charges and spins can at most differ in sign. In our case, we find that $\tilde{\rho}_{Lm}=\tilde{\rho}_{Rm}$ and $\tilde{S}^{z}_{Lm}=\tilde{S}^{z}_{Rm}$, whereas  $\tilde{S}^{x}_{Lm}=-\tilde{S}^{x}_{Rm}$, see the Appendix~\ref{App_XSpinCharge}. We confirm that for values of $m$ such that ${\rm max} \{ {\xi_{e}},{\xi_{i}} \}/a \ll m \ll N/2$, $\tilde{\rho}_m$ and $\tilde{S}^{x,z}_m$ converge to a constant values $\tilde{\rho}_{R,L}$ and  $\tilde{S}^{x,z}_{R,L}$. Without loss of generality, in what follows, we focus only on the left boundary charge and spin.

\begin{figure}[t!]
    \renewcommand\figurename{Figure} 
\centering
\includegraphics[width=\columnwidth]{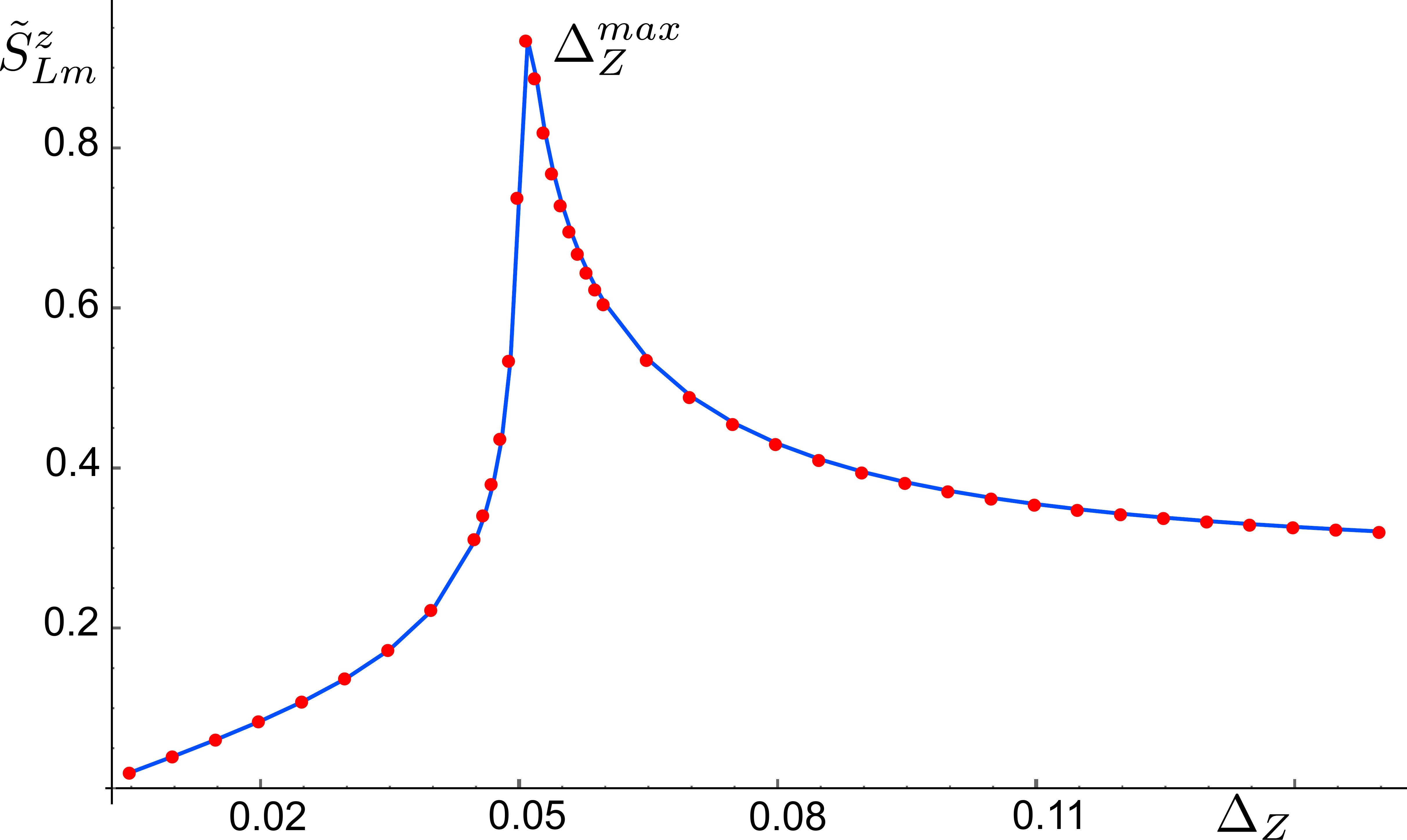}
  \caption{\label{EdgeDensPlot}The left boundary spin component along $z$ direction, $\tilde S^z_{Lm}$, [see Eq.~(\ref{EdgeDens})] as a function of the Zeeman energy $\Delta_Z$. The $\tilde S^z_{Lm}$ has a peak at $\Delta_Z^{max}$, which coincides in sufficiently long systems with the point of the topological phase transition $\Delta_Z=\Delta_{SC}$. This peak is an independent signature of the topological phase transition. The system parameters are taken to be $\alpha=0.3, \Delta_{SC}=0.05$, $\mu=0$, $N=2000$, and $m=1000$. }
\end{figure}

\section{Signature of topological phase transition}
\begin{figure}[t!]
    \renewcommand\figurename{Figure} 
\centering
\includegraphics[width=\columnwidth]{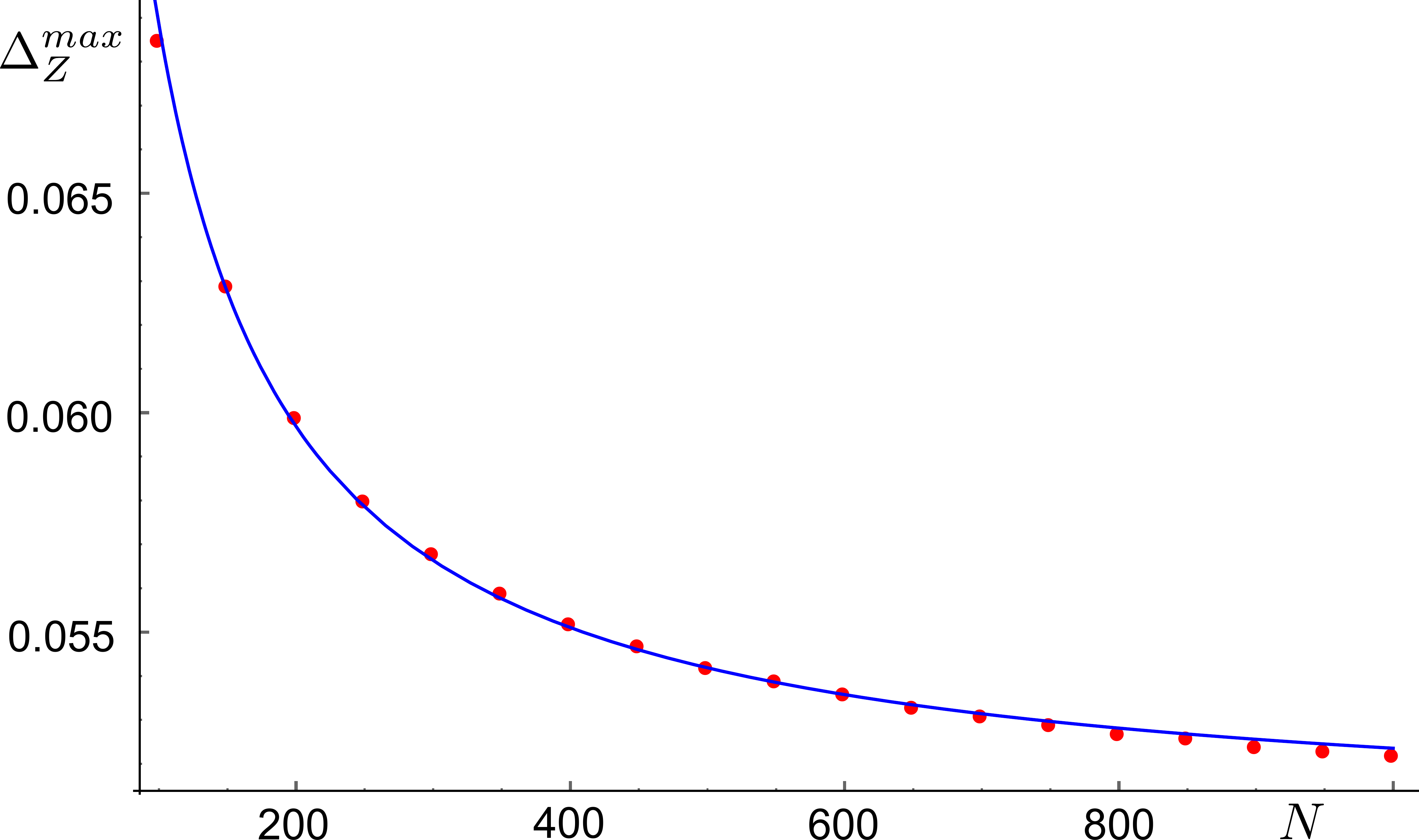}
  \caption{\label{ZDensPeakPosNPLot} The position of the peak $\Delta_Z^{max}$ in $\tilde{S}_{Lm}^{z}$ [see Eq.~(\ref{EdgeDens})] as a function of system size $N$. As the system size is increases, $\Delta_Z^{max}$ gets more and more close to the critical value $\Delta_Z^c$ at which the topological phase transition takes place. We find that the obtained numerically results (red dots) can be fitted the best with the analytical formula $(\Delta_Z^{max}-\Delta_Z^c) \propto 1/N$ (blue curve). The system parameters are the same as in Fig.~\ref{EdgeDensPlot} with $m=N/2$.}
  \end{figure}

Next, we focus on the characteristic behavior of the boundary charge and spin around the topological phase transition. First, we analyze the behaviour of the spin density along the magnetic field $S^z_j$ for various values of Zeeman gaps and all the other parameters fixed, see Fig.~\ref{ZSpinDensPlot}. As expected, $S^z_j$ is constant in the middle of the chain and saturates to $S_0^z$, however, as one approaches the end of the chain, spatial oscillations in $S^z_j$ begin to emerge. Not surprisingly, the spin polarization along the magnetic field strongly depends  on the strength of the $B$-field. The stronger  the magnetic field is, the larger is the polarization, see Fig.~\ref{ZSpinDensPlot}.  Close to the phase transition point, the oscillations in $S^z_j$ at the NW ends get more pronounced and are characterized by higher amplitudes and longer decay lengths. In order to quantify these oscillations, we calculate numerically the boundary spin and charge as defined in Eq.~(\ref{EdgeDens}).  The signature of the topological phase transition can be clearly seen in the $z$-component of the boundary spin, $S^z_{L/Rm}$, see  Fig.~\ref{EdgeDensPlot}. In the Appendix~\ref{App_XSpinCharge}, we also provide details on the boundary charge and the $S^x$-component, however, there is no signature of the topological phase transition in these quantities. In contrast to that, the $S^z_{L/Rm}$ has a pronounced peak at the value of the Zeeman energy $\Delta_Z^{max}$ that is very close to the critical value $\Delta_{Z}^c$ determined from the topological criterion. The longer the system is, the more close $\Delta_Z^{max}$ to $\Delta_{Z}^c$, see Fig.~\ref{ZDensPeakPosNPLot}. We find that $\Delta_Z^{max}$ weakly depends on the system size $N$ and approaches the critical value asymptotically as a function of $1/N$. Importantly, the value of $S^z_{L/Rm}$ does not depend on whether the MF state is occupied or not as its contribution is identically zero.
Thus, for long enough systems, the position of the peak in $S^z_{L/Rm}$ can be used as an independent signature of the topological phase transition.

\begin{figure}[t!]
\renewcommand\figurename{Figure} 
\centering
\includegraphics[width=\columnwidth]{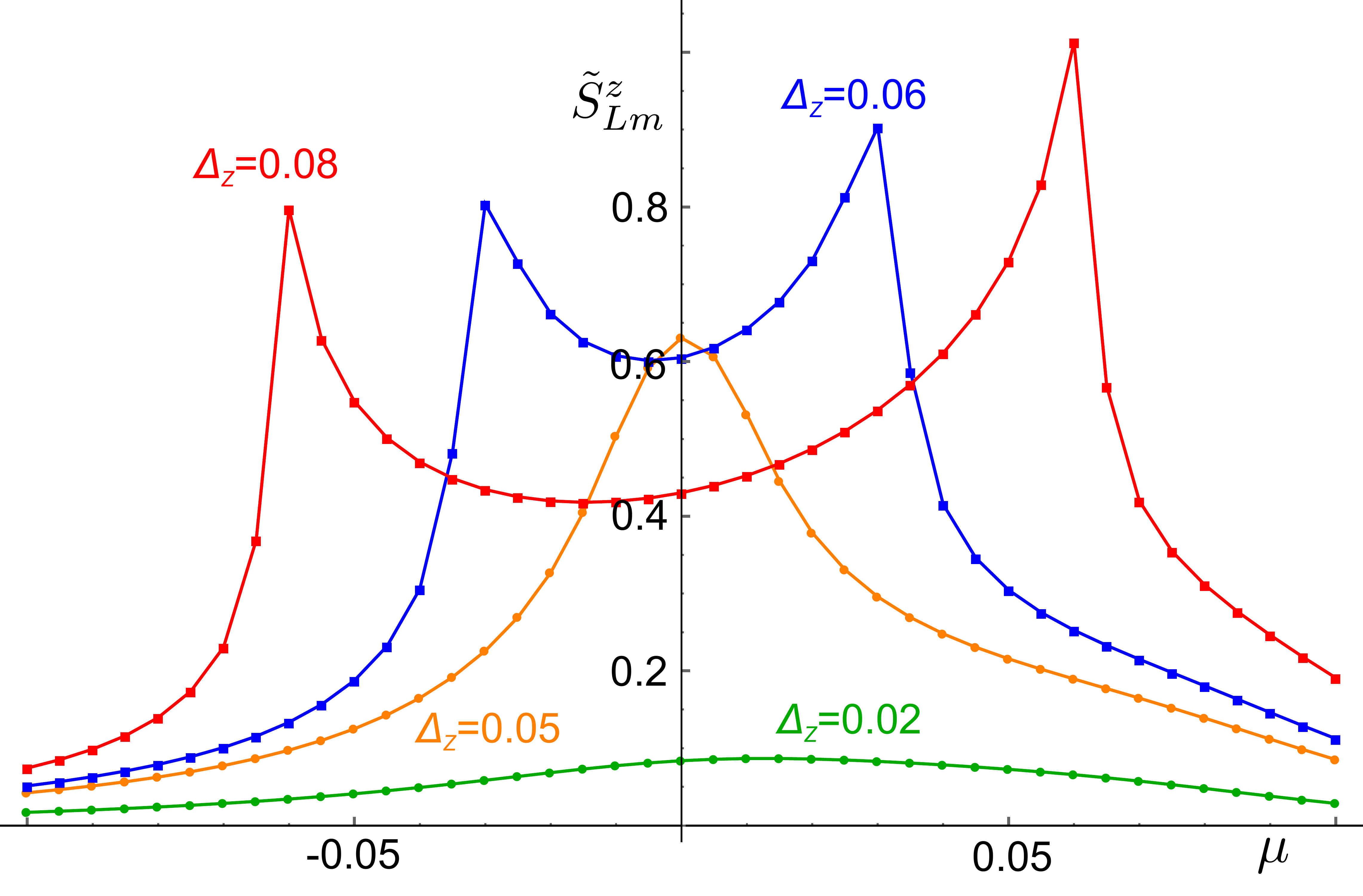}
  \caption{\label{EdgeZDensVarMuPlot} The $z$ component of the boundary spin $\tilde{S}_{Lm}^{z}$ [see Eq.~(\ref{EdgeDens})] as a function of the chemical potential $\mu$. In the topological regime, there are two peaks corresponding to two critical values of the chemical potential $\pm \mu_c$, at which the system goes through the topological phase transition. In the trivial regime such peaks are absent as the system cannot be tuned into the topological phase. Far from the transition point (green, $\Delta_Z=0.02$) there is a broad maximum in $\tilde{S}_{Lm}^{z}$ which gets more pronounced as system approaches the topological phase transition (orange, $\Delta_Z=0.05$) and develops later into a double peak structure (blue, $\Delta_Z=0.06$ and red, $\Delta_Z=0.08$) in the topological phase. The system parameters are fixed to $N=1000$, $\alpha=0.3$, $\Delta_{SC}=0.05$, and $m=500$.}
\end{figure}

It is also important to emphasize the role of the chemcical potential $\mu$. It is well known that 
the system can  also be driven between topological and trivial phases by changing $\mu$ \cite{DSarmaNanowMaj,OppenNanowMaj}. In this case, when the topological phase is reached, there are two peaks in $\tilde{S}_{Lm}^{z}$ at two critical values $\mu=\pm \mu^{c}$ with $\mu_c = \sqrt{\Delta_Z^2-\Delta_{SC}^2}$, see Fig.~\ref{EdgeZDensVarMuPlot}. Again, we see that the critical values $\pm\mu_c$ are asymptotically reached as the size of the system is increased. However, when the width of two peaks is comparable with $\mu_c$, the two peaks will merge. Thus, this criterion works  best for large values of $\Delta_Z$ and long systems. We note that one faces the same problem if the detection of the phase transition is done via zero-bias peak signatures in transport measurments. In short nanowires, the MFs of opposite ends will overlap and split away from zero energy if one is not deeply in the topological phase.

Finally, we would like to demonstrate the stability of the presented signatures against disorder and, thus, show that they are also topologically protected. For this, we add on-site disorder to our model [see Eq.~(\ref{Hamiltonian})] as well as we modify the system by adding trivial section at the NW end. Results for the both cases are presented in the Appendices~\ref{App_Disorder},~\ref{App_SCLeads}. In all configurations, the signature of the topological phase transition in the boundary spin $\tilde{S}_{Lm}^{z}$ is still fully present.

So far we have focused on signatures of the topological phase transition to be detected in the boundary spin. However, the bulk values of the spin component along the magnetic field $S^z_0$ also carry information about the topological phase transition if periodic boundary conditions are imposed, see the Appendix~\ref{App_BulkSpin} for details.  In this case, the system is translationally invariant and it does not matter at which point one computes the bulk value of the spin component $S^z_{PBC}$. The signature of the topological phase transition is still present but different. In particular, there is now a sharp discontinuity in $S^z_{PBC}$ with a jump of order $1/N$ at the point of the topological phase transition, $\Delta_Z=\Delta_Z^c$,  see the Appendix~\ref{App_BulkSpin}.  

The measurement of boundary spins will be challenging  but seems to be within reach for state-of-the-art magnetometry with NV-centers or nanoSQUIDs \cite{nanoSQUIDsWerner, LocalSpinSuscStano,NVMagnetometryScienceWrachtrup,NVMagnmetrTrifunovic,
NanSQUID,NVMagnetMalet,NanSQUID}.  We furthermore recall that it is the total integral over the spin density within the localization length that determines the spin signature of the phase transition. Thus, a resolution of the measurement device over this length scale should be sufficient and is already reached in the aformentioned magnetometric measurements. Moreover, all those techniques were already perfomed at cryogenic temperatures necessary for our proposal as one should work at temperatures that do not exceed the scale set by the bulk gap \cite{TewariTemperEff}. 
Finally, in contrast to STM measurements, these techniques are non-invasive and, thus, can be used to measure reliably the magnetic signals we propose.

\section{Conclusions}
We have identified signatures of the topological phase transition in the boundary spin component in one-dimensional topological systems. These signatures are present when tuning through the phase transition point either with the magnetic field or with the chemical potential.  Moreover, we have shown that these signatures do not rely on the presence of MFs and always occur at the boundary between topological and trivial sections of the NW. We have analyzed the finite-size effects of the boundary spin and shown that the position of the  peak converges to the value obtained analytically in the  continuum limit. These results are also stable with respect to disorder.

\begin{acknowledgments} 
We are grateful to S. Hoffman, P. Szumniak and D. Chevallier for valuable discussions.  We  acknowledge support by the Swiss National Science Foundation and the NCCR QSIT. This project has received funding from the European Union’s Horizon 2020 research and innovation program (ERC Starting Grant, grant agreement  No 757725). 
\end{acknowledgments}

\appendix

\newpage

\bigskip 

\onecolumngrid
\widetext

\begin{figure}[t!]
    \renewcommand\figurename{Figure} 
\centering
\includegraphics[width=0.8\columnwidth]{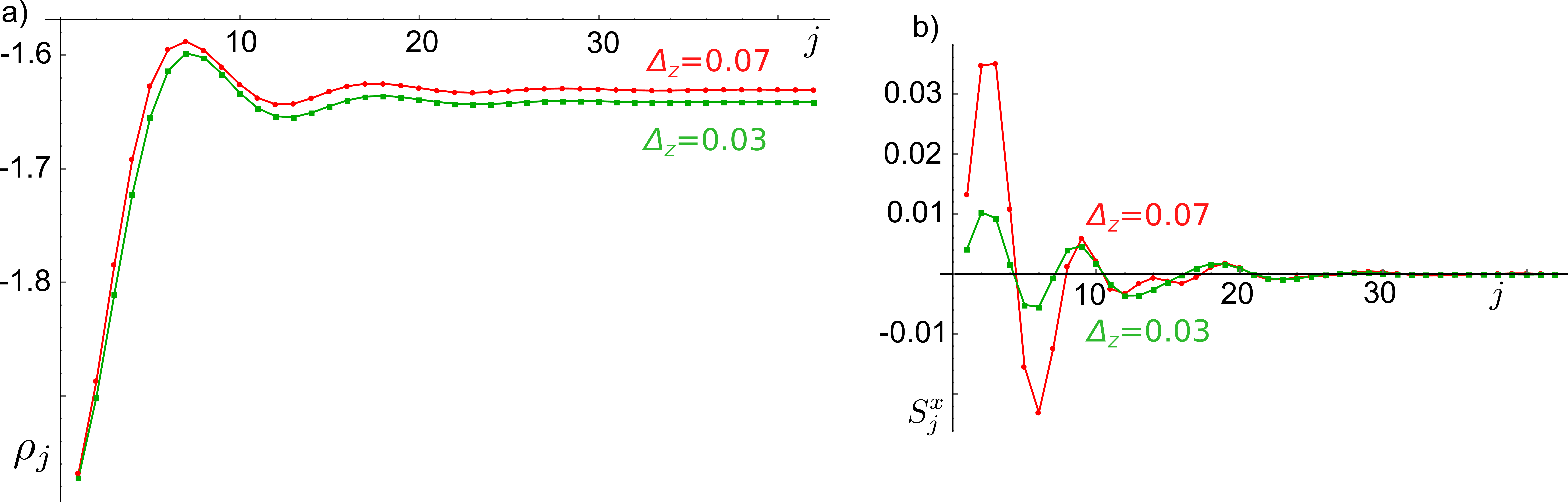}
  \caption{\label{ChargeXSpinDensPlot} (a) Charge density $\rho_j$  and (b) component of the spin density along the $x$ axis,  $S^x_j$,  as a function of site position $j$. The characteristic features are similar to those of $S^z_j$ discussed in the main text, again there are oscillations close to the NW ends in both trivial and topological phases. However, neither the amplitude of the first oscillation does not differ between the two phases [see (a)] nor the oscillations tend to cancel each other  [see (b)]. Results for the trivial (topological) phase are marked by green squares (red cycles) and correspond to $\Delta_Z=0.03$ ($\Delta_Z=0.07$). The system parameters are chosen to be $N=150$, $\mu=0$, $\alpha=0.3$, $\Delta_{SC}=0.05$. 
  }
\end{figure}

\twocolumngrid

\section{Results for boundary spin component $S^x_{Lm}$ and boundary charge $\rho_{Lm}$}
\label{App_XSpinCharge}
\begin{figure}[t!]
\renewcommand\figurename{Figure} 
\centering
\includegraphics[width=0.8\columnwidth]{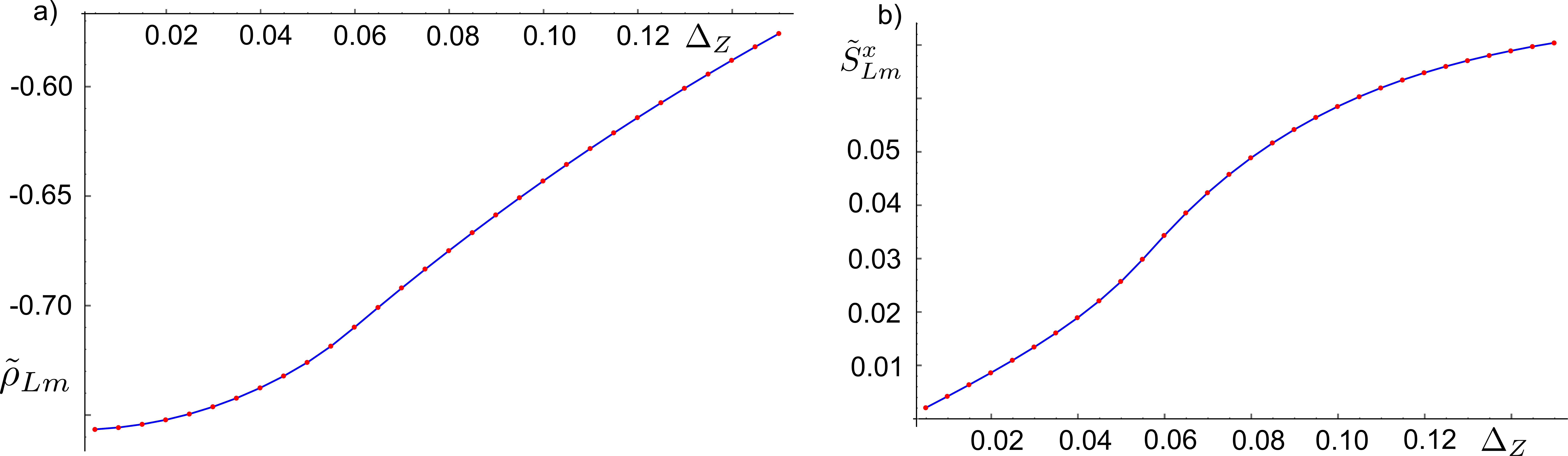}
  \caption{\label{EdgeXChargeDen}
  (a) Boundary charge  $\tilde\rho_{Lm}$  and (b) component of the boundary spin along the $x$ axis ${\tilde S}^x_{Lm}$ as a function of Zeeman energy $\Delta_Z$. There are only weak features associated with the topological phase transition. Unlike the pronounced peak  in ${\tilde S}^z_{Lm}$,  the component ${\tilde S}^x_{Lm}$ cannot reliably distinguish the trivial from topological phase. The system parameters are chosen to be $\alpha=0.3$, $\Delta_{SC}=0.05$, $\mu=0$, $N=200$, and $m=100$. }
\end{figure}

For the sake of completeness, we present here our results for the local spin density $S^x_j$ and charge density $\rho_j$, see Fig.~\ref{ChargeXSpinDensPlot}. As the SOI vector points along the $x$ direction, it is expected that $S^x_j$ in the center of the chain vanishes and moreover $\sum_{j=1}^N S_j^x=0$, which imposes that $S^x_j$ must be antisymmetric with respect to the middle of the chain. We confirm this expectation by exact numerical diagonlization. As in the case of $S^z_j$, spatial oscillations in $S^x_j$  appear close to the ends of the NW, getting more pronounced as one approaches the topological phase transition. In case of the charge density 
$\rho_j$, the characteristic behavior is very similar while in this case, as expected, the results are symmetric with respect to the middle of the wire. We also calculate the corresponding boundary charge $\tilde\rho_{Lm}$ and the boundary spin component  ${\tilde S}^x_{Lm}$ (see Fig.~\ref{EdgeXChargeDen}). However,  we do not observe any well-pronounced signatures of the topological phase transition in these quantities. For $\tilde\rho_{Lm}$, we can see a transition from almost constant to a linear dependence of $\Delta_Z$, however, this signature seems to be difficult to measure.

\onecolumngrid

\widetext

\twocolumngrid

\section{Local properties of boundary spin component  ${\tilde S}^z_{Lm}$}
\label{App_kDep}

\begin{figure}[b!]
    \renewcommand\figurename{Figure} 
\centering
\includegraphics[width=\columnwidth]{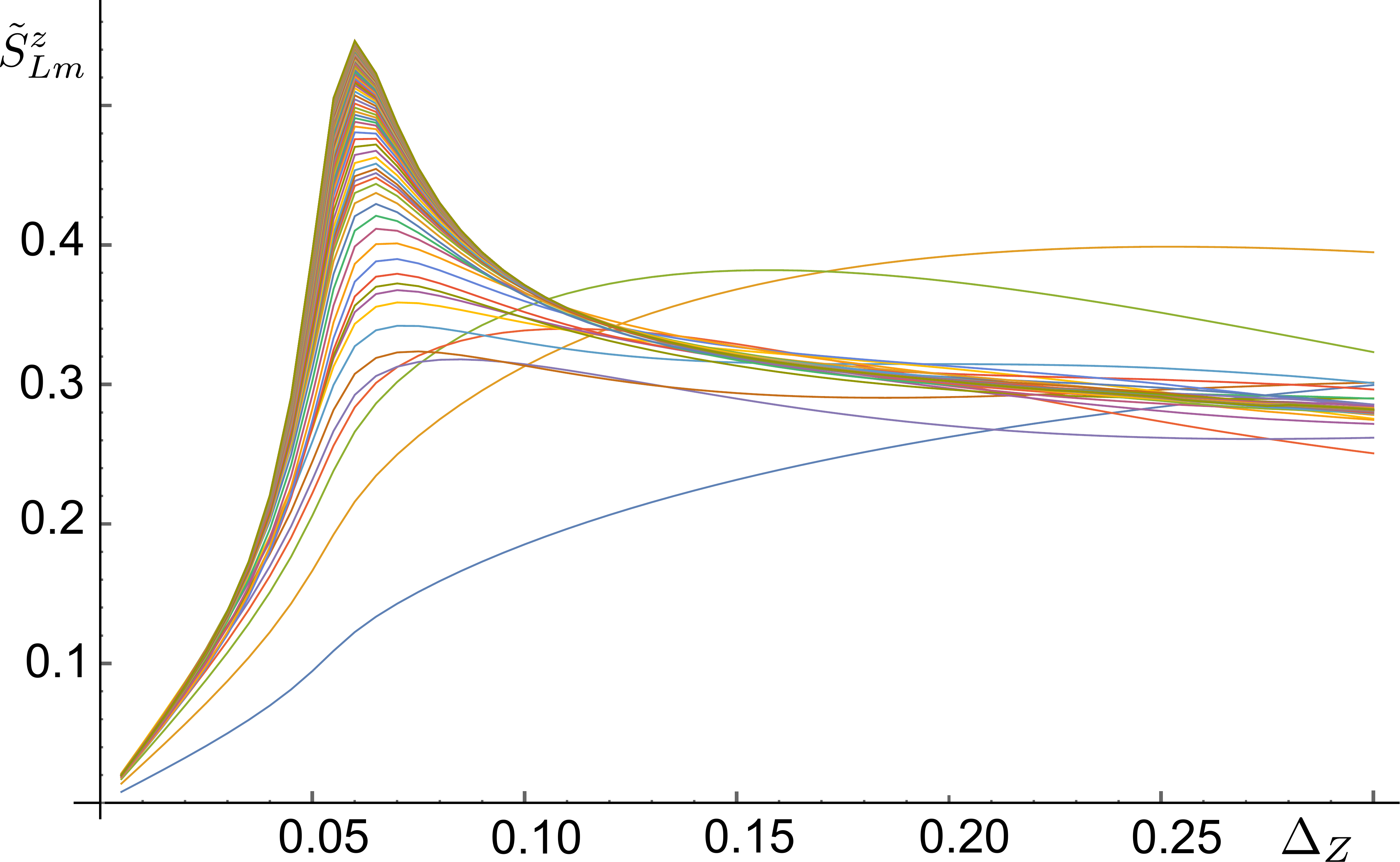}
  \caption{\label{EdgeZDensKKonvPlot} Component of the boundary spin along $z$ axis, ${\tilde S}^z_{Lm}$, [see  Eq.~(\ref{EdgeDens})] as a function of Zeeman energy $\Delta_Z$ for different cut-off values: $m=1,...,100$. Here, $m=1$ corresponds to the lowest blue curve. Other curves are ordered according to ascending $m$ at small $\Delta_Z$. The peak in ${\tilde S}^z_{Lm}$ close to $\Delta_Z^c$  is observed already for $m=\xi_{SC}/a$ ($\xi_{SC}/a=12$ for this plot), thus, the proposed signature of the topological phase transition is local with contributions coming from the occupied bulk states at the end of the NW. The system parameters are chosen to be  $N=200$, $\alpha=0.3$, $\Delta_{SC}=0.05$, and $\mu=0$. }
\end{figure}

We would like to elaborate on the question in which sense ${\tilde S}^z_{Lm}$ is a \textit{local} signature emerging only at the end of the NW. In other words, we should investigate the behavior of ${\tilde S}^z_{Lm}$ with respect to changes in $m$, see  Fig.~\ref{EdgeZDensKKonvPlot}. Far from the topological phase transition, we observe that  ${\tilde S}^z_{Lm}$  converges very quickly with increasing $m$ and is therefore a local property of the end of the NW. As we approach the transition point, values for the respective $m$'s start to differ. Nevertheless, even for $m=20$ we still observe a well-pronounced peak almost at the same $\Delta_Z$ as for $m=100$. Based on that we can conclude that  ${\tilde S}^z_{Lm}$ is a local quantity with main support at the end of the NW.

\begin{figure}[b!]
    \renewcommand\figurename{Figure} 
\centering
\includegraphics[width=\columnwidth]{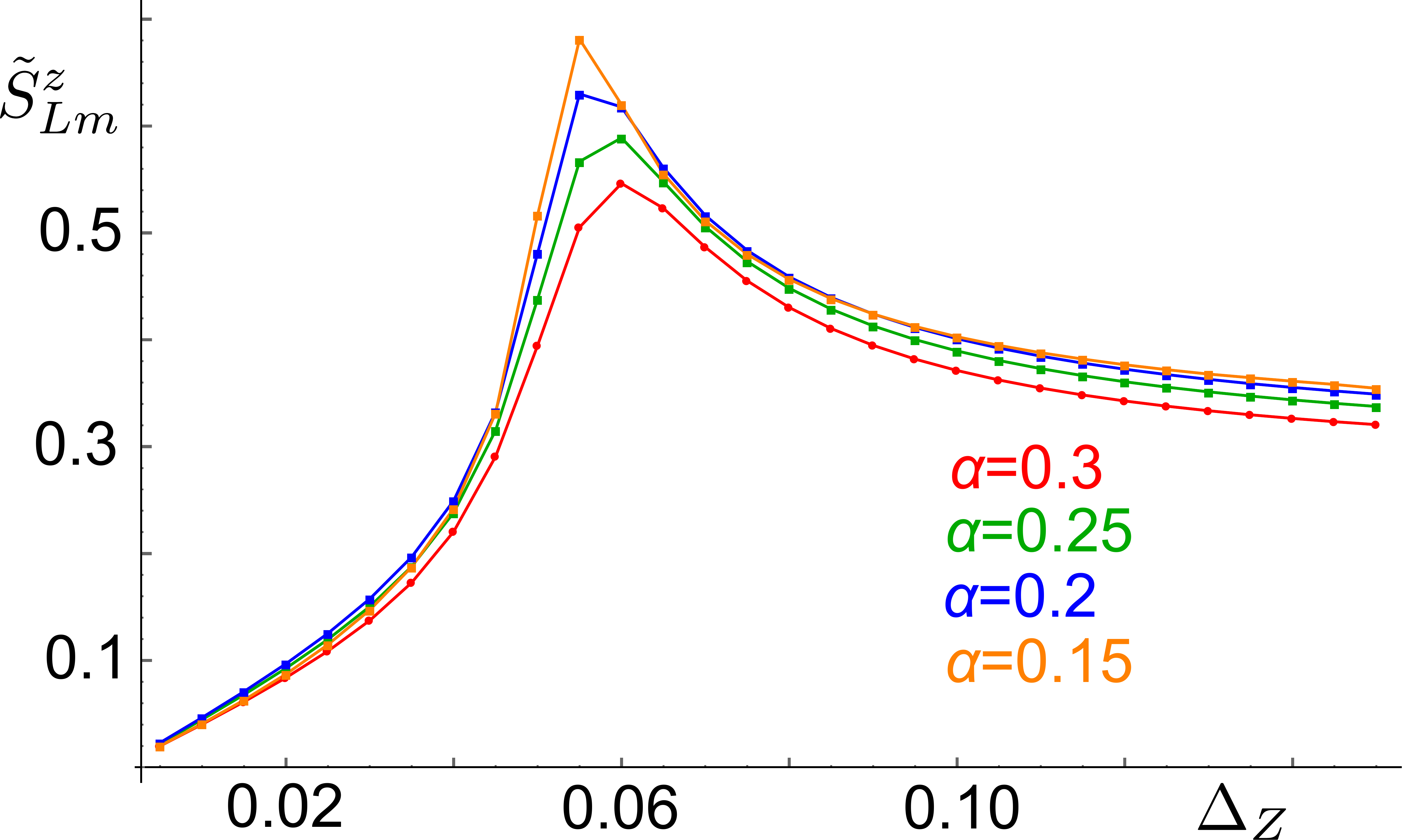}
  \caption{\label{WeakSOIPlot} Component of the boundary spin along $z$ axis, ${\tilde S}^z_{Lm}$, [see  Eq.~(\ref{EdgeDens})] as a function of Zeeman energy $\Delta_Z$ for different  values of SOI. We compare the experimentally relevant regime of strong SOI (red), $\alpha=0.3$, with intermediate SOI regimes, $\alpha=0.25$ (green) and $\alpha=0.2$ (blue), as well as with weak SOI regime, $\alpha=0.15$ (orange). The peak gets even more pronounced as one tunes from strong to weak SOI regime. The system parameters are chosen to be  $N=200$,  $\Delta_{SC}=0.05$, $m=100$ and $\mu=0$. }
\end{figure}

For completeness, we also show that the signature of the topological phase transition in ${\tilde S}^z_{Lm}$ does not crucially depend on a large value of the SOI strength, see Fig.~\ref{WeakSOIPlot}. Indeed, the peak is even more pronounced in the regime of weak SOI.

\section{Effect of on-site disorder - stability of topological phase transition signature in ${\tilde S}^z_{Lm}$}
\label{App_Disorder}

\begin{figure}[t!]
  \renewcommand\figurename{Figure} 
\centering
\includegraphics[width=0.7\columnwidth]{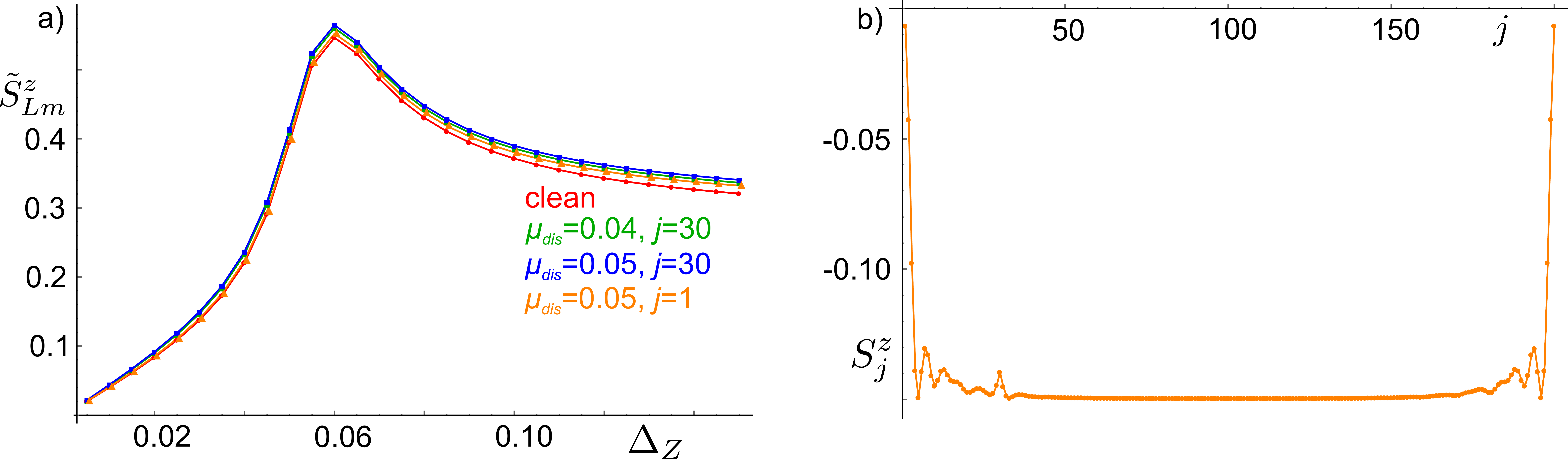}
  \caption{\label{EdgeDensDisPlot} (a) Component of the boundary spin, ${\tilde S}^z_{Lm}$, and (b) the $z$-component of the local spin density $S^z_j$ in the system with on-site potential disorder. Generally, we observe the same behavior as in the clean case, accompanied by small overall renormalization of $\tilde{S}_{j}^z$ even in the case of relatively strong disorder. We add disorder in the chemical potential $\mu_{dis}$ on site $j$, elsewhere, we keep $\mu=0$. In panel (a), ${\tilde S}^z_{Lm}$ for the clean wire (red) is compared with results for a disordered wire with $\mu_{dis}=0.04$ at site $j=30$ (green); with $\mu_{dis}=0.05$ at site $j=30$ (blue); and with  $\mu_{dis}=0.04$ at site $j=1$ (orange). If an impurity located at the first site of the NW, the boundary spin is left unchanged even for very strong values of disorder, $\mu_{dis}=0.1$. The signature of the topological phase transition is clearly not affected by disorder. In panel (b), $S^z_j$ is shown for a disordered NW with $\mu_{dis}=0.04$ at site $j=30$ ($\Delta_Z=0.07$). The presence of the impurity manifests itself as an additional local peak in $S^z_j$ accompanied by spatial oscillations. As a result, there is a local redistribution of the spin density, which does not affect the boundary spin.  If there are several impurities, their contributions average out and the system gets even more stable to disorder. 
The system parameters are chosen to be $N=200$, $\alpha=0.3$,  $\Delta_{SC}=0.05$, and $m=100$. 
  }
\end{figure}

To demonstrate that the presented signature of the topological phase transition in the boundary spin component ${\tilde S}^z_{Lm}$ is robust, we must verify that this signature persists even if the disorder is present, see Fig.~\ref{EdgeDensDisPlot}. We perform the same calculations as before, however, add fluctuations in the chemical potential. We see that, locally, disorder causes the appearance of a similar feature in the spin density as already observed at the NW ends. Namely, there is a local maximum in the spin density $S^z_j$ at the position of the impurity. The oscillations around the impurity position decay as one moves away. If there are many impurities, such effects will average out. As a result, there can be only local redistribution of the spin density $S^z_j$, which do not affect the boundary spin ${\tilde S}^z_{Lm}$. Therefore, as expected, the signature of the topological phase transition, {\it i.e.} peak in ${\tilde S}^z_{Lm}$ at $\Delta_Z^c$, is robust against local disorder. This holds also in the case of disorder as strong as the superconducting gap $\Delta_{SC}$ itself and well beyond.

\begin{figure}[h!]
  \renewcommand\figurename{Figure} 
\centering
\includegraphics[width=0.7\columnwidth]{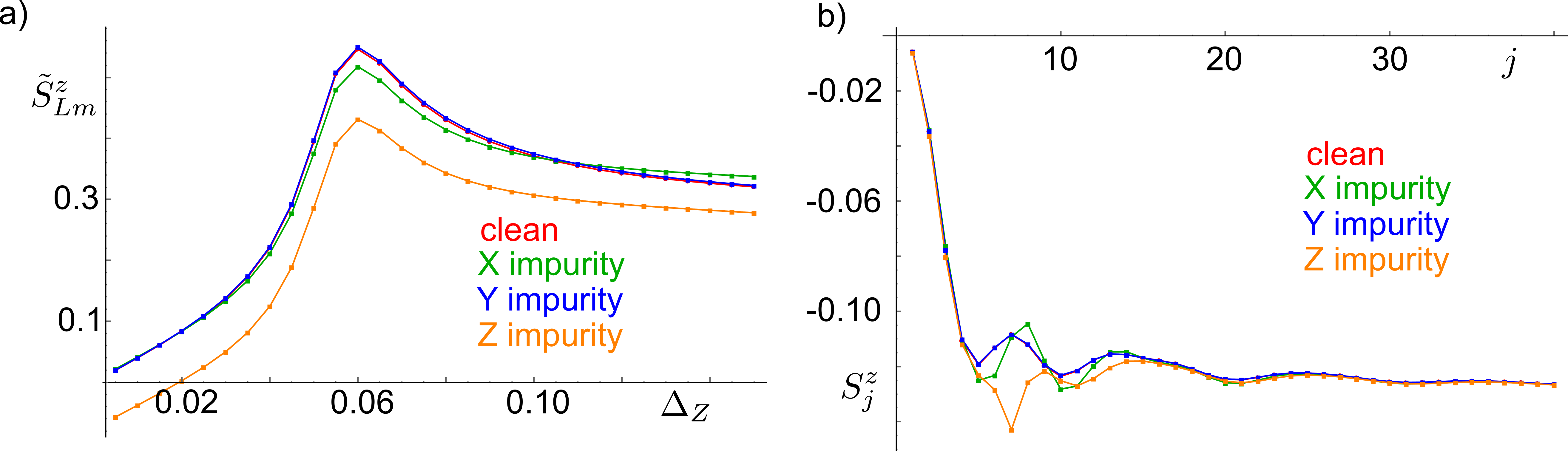}
  \caption{\label{MagDisWeakSOIPlot} 
(a) Component of the boundary spin, ${\tilde S}^z_{Lm}$, and (b) the $z$-component of the local spin density $S^z_j$ in the presence of on-site magnetic disorder. Generally, we observe the same behavior as in the clean case accompanied by small overall renormalization of ${\tilde S}^z_{Lm}$. 
 (a) Boundary spin ${\tilde S}^z_{Lm}$ for the clean wire (red) is compared with results for a disordered wire with a magnetic impurity of the strength  $J=0.05$ placed at site $j=7$ in different configurations: polarization along the $x$ direction (green) with $\theta=\pi/2$, $\phi=0$; polarization along the $y$ direction (blue) with $\theta=\pi/2$, $\phi=\pi/2$; polarization along the $z$ direction (orange) with $\theta=0$. (b) Close to the magnetic impurity, the local spin density $S^z_j$ is changed,  resulting in either an increase or decrease in the local spin polarization ($\Delta_Z=0.06$). This local redistribution of the spin density hardly affects the boundary spin and does not obscure the signature of the topological phase transition.
If there are several magnetic impurities, their contributions average out. The system parameters are chosen to be $N=200$, $\alpha=0.3$,  $\Delta_{SC}=0.05$, and $m=100$ if not specified otherwise.
}  
\end{figure}

Next, we add magnetic disorder. A magnetic impurity at site $j$ pointing in arbitrary direction defined by two spherical angles ($\theta$, $\phi$) is modeled by adding the following term to the total Hamiltonian $H$,
\begin{align}
&H_{MI,j}=J\sum_{\sigma, \sigma^{\prime}}c_{j\sigma}^{\dagger}\left(\sin\theta \cos\phi \ \sigma^{\sigma\sigma^\prime}_x+\sin\theta\sin\phi \ \sigma^{\sigma\sigma^\prime}_y\right.\nonumber\\&\hspace{130pt}+\left.\cos\theta \ \sigma^{\sigma\sigma^\prime}_z\right) c_{j\sigma^{\prime}}.
\end{align}
We repeat the same procedure as described before for potential disorder and again compare the results with the case of the clean wire, see Fig.~\ref{MagDisWeakSOIPlot}. In case of a magnetic impurity pointing in the $z$ direction along (opposite to) the direction of magnetic field, there is a dip (peak) in the local spin density. Such an effective local magnetic field sums up with the externally applied uniform field and increases (decreases) the total spin polarization, and, thus, affects the height but not the position of the peak in the boundary spin.
 In the case of the magnetic impurity pointing along the $x$ direction, there is a peak in the local spin density. This can be understood as follows: the local magnetic field polarizes spins locally along the $x$ direction, and, thus, diminishes the polarization in the $z$ direction, resulting in a local peak. In the case of a magnetic impurity pointing along the $y$ direction, there are practically no changes in the local spin density of states nor in the boundary spin. If the magnetic impurity is far away from the boundary, there is no effect on the boundary spin. In case of multiple magnetic impurities such effects average out. To conclude, magnetic disorder does not affect the signature of the topological phase transition carried by the boundary spin.

\onecolumngrid

\begin{figure}[t!]
    \renewcommand\figurename{Figure} 
\centering
\includegraphics[width=0.8 \columnwidth]{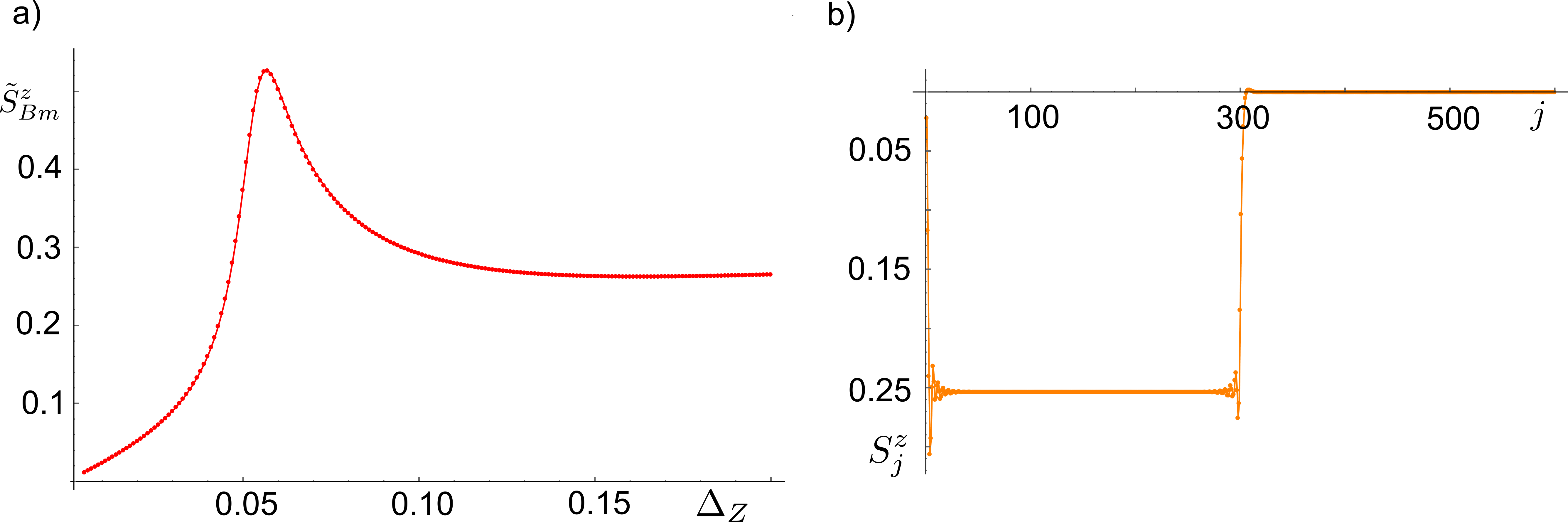}
  \caption{\label{ZDensSCAttPlot} (a) Component of the boundary spin, ${\tilde S}^z_{Bm}$, and (b) the $z$-component of the local spin density $S^z_j$. The right section of the NW corresponds to a superconducting lead, in which we fix $\alpha=\Delta_Z=0$. In contrast to that, the left section of the NW is described by $\alpha=0.3$. The chemical potential is uniform, $\mu=0$, as well as the superconducting strength $\Delta_{SC}=0.05$. In addition, $N_{<}=N_{>}=300$ and $m_{<}=m_{>}=150$.  (a) Again, there is a signature of the topological phase transition in ${\tilde S}^z_{Bm}$. (b) The boundary spin has contributions from both topological and trivial sections of the NW ($\Delta_Z=0.15$). We also note that the bulk values of spin density $S^z_j$ are different in two sections.
  }
\end{figure}

\begin{figure}[h!]
    \renewcommand\figurename{Figure} 
\centering
\includegraphics[width=0.8 \columnwidth]{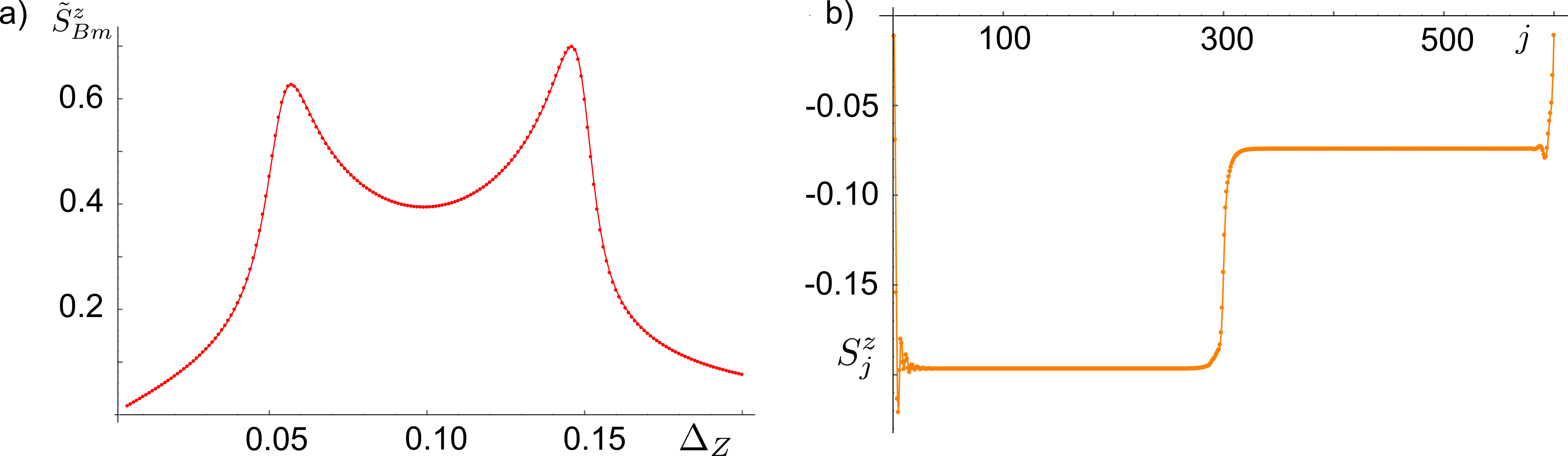}
  \caption{\label{ZDensTrivTopoAttachPlot} The same as in Fig. \ref{ZDensSCAttPlot}, however, here, both sections of the NW have the same strength of the SOI ($\alpha=0.3$) and the same strength of Zeeman energy. However, the proximity-gap is non-uniform: {\it i.e.} the right (left) section has $\Delta_{SC}=0.15$ ($\Delta_{SC}=0.05$). As a result, in panel (a), there are two peaks in  ${\tilde S}^z_{Bm}$, which corresponds to two values at which each of sections changes from the trivial to the topological phase. (b) The $z$-component of the local spin density $S^z_j$ saturates at two different values at the left and right sections, which motivates us to introduce $S_{0<}^z $ and $S_{0>}^z$ for each section separately ($\Delta_Z=0.10$).}
\end{figure}

\twocolumngrid

\section{Boundary spin ${\tilde S}^z_{Rm}$ at the boundary between topological and trivial phases}
\label{App_SCLeads}

In the main text, we have focused on the boundary spin located at the ends of the NW. Here, we show that, generally, the boundary spin is associated with the boundary between topological and trivial sections in the NW.  As a result, there is a contribution to ${\tilde S}^z_{L/Rm}$ coming from both sides of the boundary, {\it i.e.} from the topological section and from the trivial section. This means that the definitions for $\tilde S^{z}_{R/Lm}$ given by Eqs. (4) and (6) should be generalized. For the moment, let us focus on the boundary located at the site $N$ and introduce the boundary spin as
\begin{align}\label{BoundSpinGener}
&{\tilde S}^z_{Bm}=\sum^{N}_{j=N-m_<}\left( S^{z}_j-S_{0<}^z\right)-\sum^{N+m_>}_{j=N+1}\left( S^{z}_j-S^z_{0>}\right), \nonumber\\& S_{0<}^z = S^{z}_{j=[\frac{N_<}{2}]},  \ \ \ S_{0>}^z = S^{z}_{j=[N_<+\frac{N_>}{2}]}.
\end{align}
Here, the sum runs over $m_<$ ($m_>$) sites of the left (right) section of the NW consisting in total of  $N_<$ ($N_>$) sites, such that  $m=m_<+m_>$. Without loss of generality, we can assume that the left (right) section is in the topological (trivial) phase. Assuming that both sections are long enough, one determines the bulk value of the spin density as $S_{0<}^z $ and $S_{0>}^z$ for each section separately, as they are generally not the same. This can be seen clearly in Figs.~\ref{ZDensTrivTopoAttachPlot}(b) and \ref{ZDensSCAttPlot}(b), where we show how a typical spin density profile looks like in NWs divided into two sections.

We consider two scenarios. In the first scenario (see Fig.~\ref{ZDensSCAttPlot}), we attach a superconducting lead at the right end of the NW. In this lead, we assume that the Zeeman field is screened and the SOI is absent. As a result, this NW section is always in the trivial phase. Again, one observes a well-pronounced peak in ${\tilde S}^z_{Bm}$ at Zeeman energies close to the critical value $\Delta_Z^c$. In the second scenario (see Fig.~\ref{ZDensTrivTopoAttachPlot}), the right section of the NW has stronger proximity-induced superconductivity. Thus, it enters the topological phase at larger values of Zeeman energy. As a result, there are two peaks in ${\tilde S}^z_{Bm}$. The first (second) peak corresponds to a Zeeman energy  at which left (right) section of the NW becomes topological.

\section{Signatures of topological phase transition in bulk values of spin}\label{App_BulkSpin}

\begin{figure}[h!]
    \renewcommand\figurename{Figure} 
\centering
\includegraphics[width=0.8 \columnwidth]{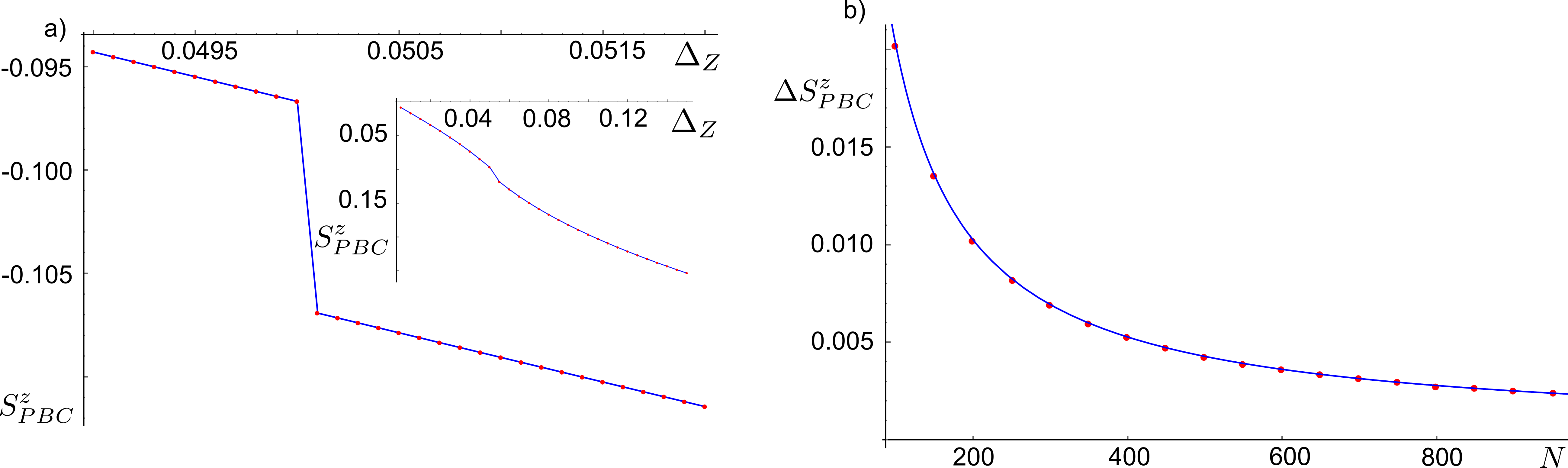}
  \caption{\label{ZSpinPBCIns} (a)  Component of spin polarization along the $z$ axis, $S^z_{PBC}$, as a function of Zeeman energy in the system with periodic boundary conditions. Away from the topological phase transition point $\Delta_Z^c$, $ S^z_{PBC}$ is a linear function of  $\Delta_Z$.  The discontinuity in spin polarization, $\Delta S^z_{PBC}$, occurs exactly at  $\Delta_Z^c$. (b) The size of the jump $\Delta S^z_{PBC}$ is inversely proportional to the system size $N$. Numerical results (red dots) are fitted by the analytical formula $\Delta S^z_{PBC} \propto 1/N$ (blue curve).
The system parameters are chosen to be $\alpha=0.3$, $\Delta_{SC}=0.05$, $\mu=0$, and $m=N/2$
 }
\end{figure}

So far we have focused on signatures of the topological phase transition to be detected in the boundary spin. However, the bulk values of the spin polarization along the magnetic field, $S^z_0$, also carries the information about the phase transition in finite-size systems. To focus on bulk properties only and to exclude any influence of boundary effects, we impose now periodic boundary conditions on the system, forming a NW ring. In this case the system is translationally invariant and it does not matter at which point one computes the bulk value of the spin polarization $S^z_{PBC}$. In finite-size systems, we always observe a sharp discontinuity in $S^z_{PBC}$ at the point of the topological phase transition, $\Delta_Z=\Delta_Z^c$,  see Fig.~\ref{ZSpinPBCIns}. In contrast to the boundary spin, this discontinuity always takes place at $\Delta_Z^c$ independent of the system size. 
However, the value of the jump $\Delta S^z_{PBC}$ in $S^z_{PBC}$  depends on the size of the system. We analyzed the value of the jump as a function of system size $N$ and conclude that it can be fitted best by an $1/N$ dependence. We note that the results of this subsection obtained for bulk states with periodic boundary conditions are closely related to the ones obtained for bulk states in Ref.~\onlinecite{SzumniakEdgeStates}. In particular, the sign reversal of the spin polarization of the bulk state with zero momentum is responsible for the jump in $S^z_{PBC}$.  In stark contrast, the features of the boundary spin $\tilde S^z_{Bm}$ are due to boundary effects and thus are of different physical origin.

\onecolumngrid

\twocolumngrid
\bibliography{referencies.bib}

\end{document}